\documentclass[prl,twocolumn,showpacs,amsmath,amssymb]{revtex4}
\usepackage[dvips]{graphicx}
\usepackage{enumerate}
\newcommand \beq{\begin{eqnarray}}
\newcommand \eeq{\end{eqnarray}}

\def\simge{\mathrel{%
       \rlap{\raise 0.511ex \hbox{$>$}}{\lower 0.511ex \hbox{$\sim$}}}}
\def\simle{\mathrel{
       \rlap{\raise 0.511ex \hbox{$<$}}{\lower 0.511ex \hbox{$\sim$}}}}

\begin{document}

\title{New Critical Point Induced by the Axial Anomaly in Dense QCD}
\author{Tetsuo Hatsuda,$^{1}$ Motoi Tachibana,$^{2}$ Naoki Yamamoto$^{1}$
and Gordon Baym$^{3}$}
\affiliation{
$^{1}$Department of Physics, University of Tokyo, Japan\\
$^{2}$Department of Physics, Saga University, Saga 840-8502, Japan\\
$^{3}$Department of Physics, University of Illinois, 1110 W. Green St.,
Urbana, Illinois 61801}

\begin{abstract}

    We study the interplay between chiral and diquark condensates within the
framework of the Ginzburg-Landau free energy, and classify possible phase
structures of two and three-flavor massless QCD.  The QCD axial anomaly acts
as an external field applied to the chiral condensate in a color
superconductor and leads to a crossover between the broken chiral symmetry and
the color superconducting phase, and, in particular, to a new critical point in
the QCD phase diagram.

\end{abstract}
\maketitle

    Quantum chromodynamics at finite temperature, $T$, and chemical potential,
$\mu$, exhibits a rich phase structure, from the hadronic Nambu-Goldstone (NG)
phase at low $T$ and $\mu$, to the quark-gluon plasma (QGP) at high $T$, and
color superconductivity (CSC) \cite{RWA} at high $\mu$.  The phase transition
from the NG phase to the QGP is being studied experimentally in
ultrarelativistic heavy ion collisions at RHIC, and will be in the future at
the LHC \cite{YHM}.  The transition from the NG to the CSC phase could also be
relevant in neutron and quark stars.

    The properties of the phases of dense matter in QCD depend on an important
interplay between two competing phenomena:  quark-antiquark pairing,
characterized by a chiral condensate $\langle \bar{q}q \rangle$, and
quark-quark pairing, characterized by a diquark condensate $\langle qq
\rangle$ \cite{NJL-model}.  We investigate here how this interplay determines
the transition between the NG and the CSC phases, via a model-independent
Ginzburg-Landau (GL) approach in terms of two order-parameter fields, the
chiral condensate $\Phi$ and diquark condensate $d$.  This phenomenon is
interesting not only in its own right, but also to other systems, e.g. the
interplay between magnetically ordered phases and metallic superconductivity
\cite{SU91}.

    We consider two simple but non-trivial cases, 3-flavor quark matter with
equal numbers of massless up ($\rm u$), down ($\rm d$), and strange ($\rm s$) quarks, and 2-flavor 
quark matter with equal numbers of massless $\rm u$ and $\rm d$ quarks.  
Chiral and color symmetries rather stringently constrain the possible couplings 
between $\Phi$ and $d$.  As we show, the $\Phi$-$d$ coupling induced by the axial 
anomaly leads to a crossover between the NG phase and the CSC phase and also to 
a new critical point in the QCD phase diagram.  The former may be relevant to the
question of continuity of the hadronic and quark matter \cite{SW99}.  The case
of two light quarks ($\rm u$ and $\rm d$) and one medium-heavy ($\rm s$) quark in
$\beta$-equilibrium with charge neutrality will be reported elsewhere.

    The GL free energy in three spatial dimensions in massless three-flavor
QCD is $\Omega = \Omega_{\chi} + \Omega_{d} + \Omega_{\chi d}$.  The
color-singlet chiral field $\Phi_{ij} \sim - \langle \bar{q}_R^j q_L^i
\rangle$ is described by the standard free-energy \cite{PW84}:
\beq
  \Omega_{\chi} &=& \frac{a_0}{2} {\rm Tr}  \  \Phi^{\dagger} \Phi
 + \frac{b_1}{4!} \left( {\rm Tr} \ \Phi^{\dagger} \Phi    \right)^2
  + \frac{b_2}{4!} {\rm Tr} \left( \Phi^{\dagger} \Phi    \right)^2 \nonumber
 \\
  & &   - \frac{c_0}{2} \left( {\rm det} \Phi + {\rm det} \Phi^{\dagger}
  \right),
 \label{eq:GL-chi}
\eeq
where ``Tr" and ``det" are taken over the flavor indices, $i$ and $j$.
Under $SU(3)_L \times SU(3)_R \times U(1)_B \times U(1)_A $ rotations, the
chiral field transforms as $\Phi \to {\rm e}^{2i \alpha_{_A}} V_{L} \Phi
V_{R}^{\dagger}$ where the phase $\alpha_{_A}$ is associated with the $U(1)_A$
rotation.  The first three terms on the right of Eq.~(\ref{eq:GL-chi}) are
invariant under this rotation; the fourth term, caused by the axial anomaly,
breaks the $U(1)_A$ symmetry down to $Z_{2N_f}=Z_6$.  We assume that $c_0 >
0$, a necessary condition for the $\eta'$ mass to obey $m_{\eta'}^2 > 0$ for
positive $\Phi$.  We also assume that the chiral phase transition is driven by
$a_0$ changing sign.  The cubic determinant term in $\Omega_{\chi}$ makes the
chiral transition with three flavors first order.

    We focus on Lorentz scalar diquarks belonging to the fundamental
representations in color and flavor space:  $\langle (q_{_L})_b^j C
(q_{_L})_c^k \rangle \sim \epsilon_{abc} \epsilon_{ijk}
[d_{_L}^{\dagger}]_{ai}$ and $\langle (q_{_R})_b^j C (q_{_R})_c^k \rangle \sim
\epsilon_{abc} \epsilon_{ijk} [d_{_R}^{\dagger}]_{ai}$ where $i, j, k$
($a,b,c$) are the flavor (color) indices, and $C$ is the charge conjugation
matrix.  Under $SU(3)_L \times SU(3)_R \times U(1)_B \times U(1)_A \times
SU(3)_C \equiv {\cal G}$, $d_{_{L,R}}$ transforms as $d_R \rightarrow e^{2i
(\alpha_B+ \alpha_A)} V_R d_R V_C^{\rm t}, \ d_L \rightarrow e^{2i (\alpha_B -
\alpha_A)} V_L d_L V_C^{\rm t}$; then $(d_R^{\ } d_L^{\dagger}) \rightarrow
e^{4i \alpha_A} V_R (d_R^{\ } d_L^{\dagger}) V_L^{\dagger}$.  The most general
form of the GL free energy of the $d$ field, to ${\cal O}(d^4)$, is
\cite{IB,GR,IMTH}:
\beq
 \label{eq:GL-d}
 \Omega_{d}
 &=&
 \alpha_0 \ {\rm Tr} [d_L^{\ } d_L^{\dagger} +d_R^{\ }  d_R^{\dagger} ]
 \nonumber \\
 &+&  \beta_{1} \left(  [ {\rm Tr}(d_L^{\ } d_L^{\dagger})]^2
                      + [ {\rm Tr}(d_R^{\ } d_R^{\dagger})]^2    \right)
 \nonumber \\
 &+& \beta_{2}  \left(  {\rm Tr} [(d_L^{\ } d_L^{\dagger})^2]
                      + {\rm Tr} [(d_R^{\ } d_R^{\dagger})^2]    \right) \\
  \nonumber
 &+& \beta_3 \ {\rm Tr} [(d_R^{\ } d_L^{\dagger})(d_L^{\ }d_R^{\dagger})]
 +  \beta_4 \ {\rm Tr} (d_L^{\ }d_L^{\dagger}){\rm Tr}(d_R^{\ }d_R^{\dagger}).
\eeq
This free energy is invariant under $\cal G$.  We assume that the normal-CSC
transition is driven by $\alpha_0$ changing sign.  Since $d_{L,R}$ carries
baryon number, ${\rm det}\ d_{L,R}$ terms are not allowed, unlike for $\Phi$.

    The interaction free energy of the chiral and diquark fields is, to fourth
order,
\beq
 \label{eq:GL-coup}
 \Omega_{\chi d}
 &= & \gamma_{1} \ {\rm Tr} [  (d_R^{\ } d_L^{\dagger})\Phi
                  + (d_L^{\ } d_R^{\dagger})\Phi^{\dagger}]     \nonumber \\
 &&+\lambda_{1} \ {\rm Tr} [(d_L^{\ } d_L^{\dagger})\Phi \Phi^{\dagger}
                  +(d_R^{\ } d_R^{\dagger})\Phi^{\dagger}\Phi ]  \nonumber \\
&&+\lambda_{2} \ {\rm Tr} [d_L^{\ } d_L^{\dagger} + d_R^{\ } d_R^{\dagger}]
                    \cdot {\rm Tr} [\Phi^{\dagger}\Phi ] \\ \nonumber
&&+  \lambda_{3} \left( {\rm det} \Phi \cdot
                  {\rm Tr}[(d_L^{\ } d_R^{\dagger}) \Phi^{-1}] + h.c. \right) .
\eeq
The triple boson coupling $\sim \gamma_1$, which breaks the $U(1)_A$
symmetry down to $Z_6$, originates from axial anomaly.  The remaining terms
are fully invariant under $\cal G$ \cite{SS00}.  Equations
(\ref{eq:GL-chi})-(\ref{eq:GL-coup}) constitute the most general form of the GL
free energy under the conditions that the phase transition is not strongly
first order and that the condensed phases are spatially homogeneous.

    We assume, for three flavors in the chiral limit, a flavor symmetric
chiral condensate, $\Phi ={\rm diag}(\sigma, \sigma,\sigma)$, and a
color-flavor-locked (CFL) diquark condensate, $d_L=-d_R={\rm diag}(d,d,d)$, in
which all flavors contribute equally to the $J^P=0^+$ pairing.  Then the GL
free-energy, the sum of Eqs.  (\ref{eq:GL-chi})-(\ref{eq:GL-coup}), reduces to
\beq
 \label{eq:nf3-model}
 \Omega_{3F}
 &=& \left( \frac{a}{2} \sigma^2 - \frac{c}{3} \sigma^3 + \frac{b}{4}\sigma^4
 \right)
 + \left( \frac{\alpha}{2} d^2 + \frac{\beta}{4} d^4 \right) \nonumber \\
 & & - {\gamma} d^2 \sigma +  {\lambda} d^2 \sigma^2.
\eeq

    Because the cubic $d^2 \sigma$ and $\sigma^3$ terms in
Eq.~(\ref{eq:nf3-model}) both arise from the axial anomaly, $\gamma$ and $c$
are related microscopically.  Indeed, it is straightforward to show from the
instanton-induced six-fermion interaction, $\det_{i,j}(\bar{q}_R^j q_L^i)$,
that $\gamma$ has the same sign and the same order of magnitude as $c$
\cite{TS,BHTY-2}.  Positive $\gamma$ (attraction) favors coexistence, $\sigma
\neq0, d \neq 0$.  Also, the $d^2 \sigma$-term acts to break chiral symmetry
explicitly, implying that $\sigma \neq 0$ may be realized for all baryon
densities, as discussed below.

    As we can show from microscopic calculations in weak coupling QCD and in
the NJL model, $\lambda>0$ and $\beta>0$.  A non-vanishing $\sigma$ plays the
role of an effective mass for the quark field, reducing the density of states
at the Fermi surface, and the pairing energy \cite{IMTH}, an effect
represented $\lambda d^2 \sigma^2 >0$.  Furthermore, we find $\lambda/\beta
\sim \ln(\Lambda/T_d)/(\mu/T_d)^2$, which is rather small for reasonable
values of $\mu$, $T_d$ (the critical temperature of the color
superconductivity without the $\sigma$-$d$ coupling), and $\Lambda$ ($\sim
\mu$ for weak coupling QCD, and in the NJL model, $\sim$ the spatial momentum
cutoff).  We consider here the first order chiral transition driven by $c$
with positive $b$; the case with negative $b$ in three flavors does not change
the results qualitatively \cite{BHTY-2}.  On the basis of the above
discussion, we focus on the case $\gamma >0$, $\lambda \ge 0$ with $b >0$.

    In principle, the system can have four possible phases:  normal (NOR) with
$\sigma = d =0$, CSC ($\sigma$=0, $d\neq$ 0), NG ($\sigma\neq$ 0,
$d$= 0), and coexistence (COE) ($\sigma\neq$ 0, $d \neq$ 0).  We locate
the phase boundaries and the order of the phase transitions by comparing the
free energies, $\Omega^{\rm (NOR)}(0,0)$, $\Omega^{\rm (CSC)}(0,d)$,
$\Omega^{\rm (NG)}(\sigma,0)$, and $\Omega^{\rm (COE)}(\sigma,d)$.

    In the COE phase, it is useful to analyze the free energy in terms of the
single variable, $\sigma(d)$, or $d(\sigma)$, obtained by solving the
stationarity condition, $\partial \Omega^{\rm (COE)}(\sigma,d)/\partial d =0$.
\begin{figure}[t]
\begin{center}
\includegraphics[width=4.5cm]{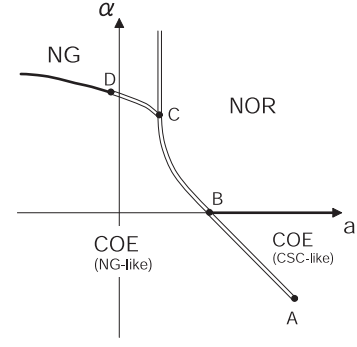}
\end{center}
\vspace{-0.5cm}
\caption{Phase structure in the three-flavor system with $\gamma >0$ and
$\lambda=0$.  The phase boundaries with a first (second) order transition is
denoted by a double (single) line.
}
\label{fig:3F}
\end{figure}

    In the absence of $\sigma$-$d$ coupling ($\gamma=\lambda=0$), the four
phases defined above are separated by $\alpha=0$ (a line of second
order transitions) and by $a=2c^2/(9b)$ (a line of first order phase
transitions).  But with an {\it attractive} coupling, $\gamma > 0$ with
$\lambda=0$, the phase structure undergoes four major modifications, as shown
in Fig.\ref{fig:3F}, where the first and second order phase boundaries are
shown by double and single lines respectively:
    (i) The area of the COE phase grows, since the $d^2 \sigma$ term
lowers the free energy if both $\sigma$ and $d$ are nonvanishing.
    (ii) The first order line between the CSC-like and the NG-like COE,
originally located at $a=2c^2/(9b)$, terminates at a critical point, A
\cite{NJL1-model}.  This behavior is expected since $d^2 \sigma$ acts as an
external field for $\sigma$, washing out the first order phase
transition for sufficiently large $\gamma$ or $d$.  The NG-like COE phase in
Fig.~\ref{fig:3F} has larger $\sigma$ than the CSC-like COE phase across the
boundary BA.
    (iii) The second order boundary originally at $\alpha=0$ splits in
two, a line going to the right from the critical end point B and a line going
to the left from the point C. Since $\sigma$ changes discontinuously across
the first order boundary CB, the $d^2 \sigma$ term, which acts as a mass term
for $d^2$, leads to different critical temperatures for diquark condensation
to the NG-like COE and CSC-like COE.
    (iv) For $\gamma >(\sqrt{\beta/b}) c/3$, a tri-critical point D appears on
the phase boundary between NG and COE.  Then the point C, otherwise a
critical end point, becomes a triple point.

\begin{table}[b]
    \caption{For three flavors, with $\gamma >0$ and $\lambda=0$: locations of
A, the critical point, B, the critical end point, C, the triple point, and D,
the tri-critical point, in Fig.~\ref{fig:3F}:  The point D appears for $\gamma
> (\sqrt{\beta/b}) c/3$.  For two flavors with $b < 0$ and $\lambda >
0$: locations of E and F in Fig.~\ref{fig:2Fn}. Here $c' \equiv
\sqrt{c^2+16b\gamma^2/\beta}$ and $b'\equiv b- 4\lambda^2/\beta$.  }
\begin{tabular}{|c|c|c|}
\hline
    & $a$ & $\alpha$  \\ \hline \hline
A   &  ${c^2}/(3b)+{2\gamma^2}/{\beta}$
    & $- {\beta c^3}/{(27 \gamma b^2)} $     \\
B   &  ${2c^2}/{(9b)}+{2\gamma^2}/{\beta}$ &  $0$         \\
C   &  ${2c^2}/{(9b)}$ &
       $\sqrt{{\beta}/{b}}
        \left(
        {c}/{(3\sqrt{b})}+ {\gamma}/{\sqrt{\beta}} \right)^2 $     \\
D   &  $    \left( c+c' \right) {c}/{(8b)}  - {\gamma^2}/{\beta}$ &
       $  \left(  c+c'  \right) {\gamma}/{(2b)} $       \\
\hline
 E &  ${3b^2}/{(16f)}$ & 0 \\
  F &  ${3b'} (b' + 16 {\lambda^2}/{\beta})/(16f) $ &
   ${3\lambda b'}/{(2f)}$ \\
\hline
\end{tabular}
 \end{table}

    Table I locates the characteristic points, A-D, in the ($a,\alpha$) plane.
To show explicitly how the critical point A appears, we derive an {\it
effective} free-energy $\Omega_{3F}[\sigma,d(\sigma)]$, using the stationarity
condition, $d^2 = 2(\gamma\sigma-\alpha/2)/\beta$:
\beq
 \label{eq:effective-sigma-energy}
 \Omega_{3F}[\sigma,d(\sigma)]
 = -\frac{\alpha^{2}}{4\beta} +\alpha^{*}\sigma+
 \frac{a^{*}}{2} \sigma^2 - \frac{c}{3} \sigma^3 + \frac{b}{4}\sigma^4,
\eeq
with $\alpha^*\equiv \alpha\gamma/\beta$ and $a^* \equiv
a-2\gamma^2/\beta$.  Equation~(\ref{eq:effective-sigma-energy}) is valid for
$\sigma \ge \alpha/(2\gamma)$.  We eliminate the $\sigma^3$ term in
Eq.~(\ref{eq:effective-sigma-energy}), by introducing $\tau =
\sigma - c/(3b)$.  Then the system becomes equivalent to an Ising ferromagnet in
an external magnetic field.  The point A in Fig.\ref{fig:3F} corresponds to
the second order critical point of this magnetic system.

    The $\lambda d^2 \sigma^2$ term, with positive $\lambda$, modifies the
coefficients $a^*$, $c$ and $b$ in Eq.~(\ref{eq:effective-sigma-energy}) as
$a^* \to a^*- 2\alpha \lambda/\beta$, $c \to c - 6\gamma \lambda/\beta$ and $b
\to b- 4 \lambda^2/\beta$.  Therefore, for $\lambda / \beta \ll 1$, as
suggested microscopically, the $\lambda$ term does not qualitatively change
the phase diagram in Fig.\ref{fig:3F}.

    Let us discuss the axial anomaly-driven crossover from the point of view
of chiral symmetry.  The CSC phase with a CFL structure ($d_L^{\
}d_R^{\dagger} =- {\rm diag}(d^2,d^2,d^2)\neq 0$) breaks chiral symmetry but
preserves the $Z_4$ discrete subgroup of $U(1)_A$.  On the other hand, in the
COE phase, $d_L^{\ }d_R^{\dagger} =-{\rm diag}(d^2,d^2,d^2)\neq 0$ and $\Phi
={\rm diag}(\sigma,\sigma,\sigma)$ lead to chiral symmetry breaking,
preserving only $Z_2$.  The symmetry breaking pattern is different in the two
phases.  However, the $\gamma_1$-term in Eq.~(\ref{eq:GL-coup}) has $Z_6$
symmetry which contains $Z_2$ (but not $Z_4$) as a subgroup.  Therefore, once
the axial anomaly is present, the NG-like and CSC-like COE phases cannot be
distinguished by symmetry and can be continuously connected.  The NG and COE
phases differ in the realization of $U(1)_B$ symmetry, and therefore their
boundary is not smoothed out.

    We turn now to the massless two-flavor system (with infinite $\rm s$ quark
mass).  In this case, all chiral and diquark condensates with an $\rm s$ quark
are suppressed; we write $\Phi ={\rm diag}(\sigma,\sigma,0)$ and
$d_L=-d_R={\rm diag}(0,0,d)$.  The latter is the two flavor color
superconductivity, 2SC state.  Due to this color-flavor structure, the cubic
terms in $\sigma$ and $d$ are identically zero, and the model reduces to:
\beq
\label{eq:nf2-model}
 \! \! \! \Omega_{2F} =
  \left( \frac{a}{2} \sigma^2 + \frac{b}{4} \sigma^4
 + \frac{f}{6} \sigma^6 \right)
 +\left(  \frac{\alpha}{2} {d}^2 + \frac{\beta}{4} {d}^4 \right)
 + {\lambda} {d}^2 \sigma^2.
\eeq
Since in two flavor QCD at finite $T$ and $\mu$, a tricritical point
($a=b=0$) may exist at which the second order transition for $b>0$ turns into
a first order transition for $b<0$ \cite{AY89}, we introduce a $\sigma^6$ term
with positive coefficient $f$ as the minimal extension of the model to
stabilize the system.

\begin{figure}[t]
\begin{center}
\includegraphics[width=7.5cm]{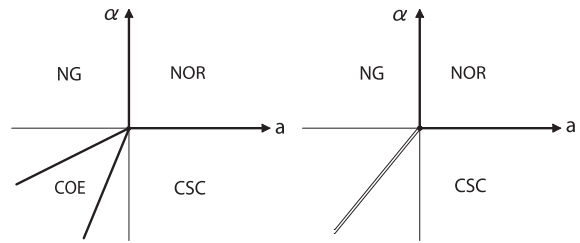}
\end{center}
\vspace{-0.5cm}
\caption{Phase strucuture in the two-flavor system for $b>0$ and $\lambda
> 0$.  Left:  Case with the tetra-critical point ($\frac{1}{2}\sqrt{b\beta} >
\lambda > 0$).  The second order line between NG and COE (CSC and COE) is
characterized by $\alpha = 2a\lambda/b$ ($\alpha = a\beta/(2\lambda)$).  Right:
Case with the bi-critical point ($\lambda > \frac{1}{2}\sqrt{b\beta}$).  The
first order line between NG and CSC is characterized by $\alpha =
a\sqrt{\beta/b}$.}
\label{fig:2Fp}
\end{figure}

    We first consider $b >0$ in Eq.~(\ref{eq:nf2-model}), neglecting the
$\sigma^6$ term in finding the qualitative phase structure \cite{CL}.  For
$\lambda=0$, the boundaries of the four phases are characterized by second
order lines at $\alpha=0$ and $a=0$ with a tetra-critical point at
$\alpha=a=0$.  With the {\it repulsive} $d^2 \sigma^2$ term ($\lambda > 0$),
the area of the coexistence phase decreases, as shown on the left of
Fig.~\ref{fig:2Fp}.  For $ \lambda > \frac{1}{2}\sqrt{b\beta}$, the
coexistence region disappears, a first order interface between CSC and NG
appears at $\alpha = a\sqrt{\beta/b}$, and $a=\alpha=0$ becomes a bi-critical
point, as shown in the right of Fig.\ref{fig:2Fp}.

\begin{figure}[t]
\begin{center}
\includegraphics[width=4cm]{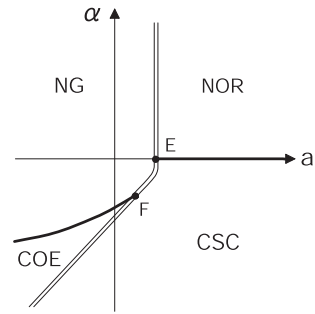}
\end{center}
\vspace{-0.5cm}
    \caption{Phase strucuture of the two-flavor system with $b<0$ and $\lambda
> 0$.  The second order line between NG and COE is $\alpha=
\lambda(b-\sqrt{b^2-4fa})/f$.  The first order line between CSC and COE is at
$\alpha=\frac{\beta}{2\lambda} [a- 3(b- 4\lambda^2/\beta)^2/(16f)]$.  The first
order line between NG and CSC is at $\alpha^2 = \beta[6bfa +
(b^2-4fa)^{3/2}-b^3]/(6f^2)$.}
\label{fig:2Fn}
\end{figure}

    Next we consider $b <0$; here the $\sigma^6$ term plays an essential role.
For $\lambda=0$, the four phases are separated by a second order line at
$\alpha=0$ and a first order line at $a=3b^2/(16f)$.  With a {\it repulsive}
$d^2\sigma^2$ term ($\lambda > 0$) the coexistence phase shrinks and gradually
fades away as $\lambda \to \infty$.  Moreover, a new first order line between
NG and CSC appears and grows as $\lambda$ increases.  This situation is shown
in Fig.~\ref{fig:2Fn}.  The locations of the critical end points E and F are
given in Table I. Such a phase structure was previously noted in
\cite{RM-model} which used the random matrix model and in \cite{NJL2-model}
which used the NJL model; our model-independent analysis is consistent with
these results.

\begin{figure}[h]
\begin{center}
\includegraphics[width=5.5cm]{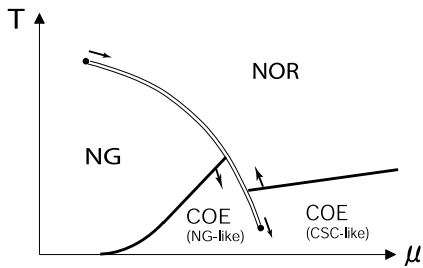}
\end{center}
\caption{Schematic phase structure with two light quarks (up and down) and
a medium heavy quark (strange).  The arrows show how the critical points and
the phase boundaries move as the strange-quark mass increases toward the
two-flavor limit.}
\label{fig:32F}
\end{figure}

    The mapping of the phase diagrams in the ($a,\alpha$) plane to the
($T,\mu$) plane is a dynamical question which we cannot address within the GL
theory.  Nevertheless, in Fig.~\ref{fig:32F}, we draw a {\em speculative}
phase structure with two light quarks and a medium-heavy quark.  There are two
critical points in the figure:  one near the vertical axis driven by finite
$\mu$ (as seen in the two-flavor case), and a new critical point near the
horizontal axis driven by the axial anomaly (as seen in the three-flavor
case).  With decreasing strange quark mass, $m_s$, the higher critical point
approaches the vertical axis, while as $m_s$ increases the lower critical
point approaches horizontal axis.  The chiral transition is a crossover in the
direction of both high $T$ and high $\mu$.  Whether this scenario is realized
or not should be eventually checked by first principles QCD simulations.

    In summary, we find that the QCD axial anomaly acts as an external field
for the chiral condensate under the influence of the diquark condensate.  The
first order chiral transition is changed to a crossover for a large diquark
condensate, and a new critical point driven by the axial anomaly emerges in
the QCD phase diagram.  Precise location this point is a future task for
phenomenological models and lattice QCD simulations.  Our schematic phase
diagram would be made more realistic by including effects, such as finite
quark masses, charge neutrality, $\beta$ equilibrium, and thermal gluon
fluctuations \cite{IMTH,MIHB}.  Open questions include whether the new
critical point would survive in an inhomogeneous Fulde-Ferrell-Larkin-Ovchinnikov 
(FFLO) state, and how the COE phase at low $T$ and $\mu$, Fig.\ref{fig:32F}, 
is affected by quark confinement.

\begin{acknowledgments}

    We thank S. Digal, K. Itakura, T. Matsuura, and K. Fukushima for helpful
comments.  This research was supported in part by the Grants-in-Aid of the
Japanese Ministry of Education, Culture, Sports, Science, and Technology
(No.~15540254), and in part by NSF Grant PHY 03-55014.  Author GB thanks the
University of Tokyo for kind hospitality and, in particular, the CNS for
support.

\end{acknowledgments}


\begin{thebibliography}{99}
\bibitem{NJL1-model} A similar critical point (CP) at low temperature has
been derived in \cite{NJL2-model} using the two-flavor NJL model with scalar
and vector type four-fermion interactions.  Since the axial anomaly does not
produce a triple boson coupling in two flavors, the origin of their CP is not
related to ours.
\end{thebibliography}
\end{document}